\documentclass[twoside,twocolumn,9pt]{article}
\usepackage{extsizes}
\usepackage[super,sort&compress,comma]{natbib} 
\usepackage[version=3]{mhchem}
\usepackage[left=1.5cm, right=1.5cm, top=1.785cm, bottom=2.0cm]{geometry}
\usepackage{balance}
\usepackage{mathptmx}
\usepackage{sectsty}
\usepackage{graphicx} 
\usepackage{lastpage}
\usepackage[format=plain,justification=justified,singlelinecheck=false,font={stretch=1.125,small,sf},labelfont=bf,labelsep=space]{caption}
\usepackage{float}
\usepackage{fancyhdr}
\usepackage{fnpos}
\usepackage[english]{babel}
\addto{\captionsenglish}{%
  
}
\usepackage{array}
\usepackage{droidsans}
\usepackage{charter}
\usepackage[T1]{fontenc}
\usepackage[usenames,dvipsnames]{xcolor}
\usepackage{setspace}
\usepackage[compact]{titlesec}
\usepackage{hyperref}
\usepackage{eucal}
\usepackage{amsmath}
\usepackage{amssymb}

\usepackage{epstopdf}%This line makes .eps figures into .pdf - please comment out if not required.

\definecolor{cream}{RGB}{222,217,201}

\begin{document}

\pagestyle{fancy}
\thispagestyle{plain}
\fancypagestyle{plain}{
%%%HEADER%%%
\renewcommand{\headrulewidth}{0pt}
}
%%%END OF HEADER%%%

%%%PAGE SETUP - Please do not change any commands within this section%%%
\makeFNbottom
\makeatletter
\renewcommand\LARGE{\@setfontsize\LARGE{15pt}{17}}
\renewcommand\Large{\@setfontsize\Large{12pt}{14}}
\renewcommand\large{\@setfontsize\large{10pt}{12}}
\renewcommand\footnotesize{\@setfontsize\footnotesize{7pt}{10}}
\makeatother

\renewcommand{\thefootnote}{\fnsymbol{footnote}}
\renewcommand\footnoterule{\vspace*{1pt}% 
\color{cream}\hrule width 3.5in height 0.4pt \color{black}\vspace*{5pt}} 
\setcounter{secnumdepth}{5}

\makeatletter 
\renewcommand\@biblabel[1]{#1}            
\renewcommand\@makefntext[1]% 
{\noindent\makebox[0pt][r]{\@thefnmark\,}#1}
\makeatother 
\renewcommand{\figurename}{\small{Fig.}~}
\sectionfont{\sffamily\Large}
\subsectionfont{\normalsize}
\subsubsectionfont{\bf}
\setstretch{1.125} %In particular, please do not alter this line.
\setlength{\skip\footins}{0.8cm}
\setlength{\footnotesep}{0.25cm}
\setlength{\jot}{10pt}
\titlespacing*{\section}{0pt}{4pt}{4pt}
\titlespacing*{\subsection}{0pt}{15pt}{1pt}
%%%END OF PAGE SETUP%%%

%%%FOOTER%%%
\fancyfoot{}
\fancyfoot[LO,RE]{\vspace{-7.1pt}\includegraphics[height=9pt]{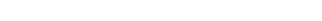}}
\fancyfoot[CO]{\vspace{-7.1pt}\hspace{13.2cm}\includegraphics{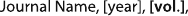}}
\fancyfoot[CE]{\vspace{-7.2pt}\hspace{-14.2cm}\includegraphics{RF}}
\fancyfoot[RO]{\footnotesize{\sffamily{1--\pageref{LastPage} ~\textbar  \hspace{2pt}\thepage}}}
\fancyfoot[LE]{\footnotesize{\sffamily{\thepage~\textbar\hspace{3.45cm} 1--\pageref{LastPage}}}}
\fancyhead{}
\renewcommand{\headrulewidth}{0pt} 
\renewcommand{\footrulewidth}{0pt}
\setlength{\arrayrulewidth}{1pt}
\setlength{\columnsep}{6.5mm}
\setlength\bibsep{1pt}
%%%END OF FOOTER%%%

%%%FIGURE SETUP - please do not change any commands within this section%%%
\makeatletter 
\newlength{\figrulesep} 
\setlength{\figrulesep}{0.5\textfloatsep} 

\newcommand{\topfigrule}{\vspace*{-1pt}% 
\noindent{\color{cream}\rule[-\figrulesep]{\columnwidth}{1.5pt}} }

\newcommand{\botfigrule}{\vspace*{-2pt}% 
\noindent{\color{cream}\rule[\figrulesep]{\columnwidth}{1.5pt}} }

\newcommand{\dblfigrule}{\vspace*{-1pt}% 
\noindent{\color{cream}\rule[-\figrulesep]{\textwidth}{1.5pt}} }

\makeatother
%%%END OF FIGURE SETUP%%%

%%%TITLE, AUTHORS AND ABSTRACT%%%
\twocolumn[
  \begin{@twocolumnfalse}
{\includegraphics[height=30pt]{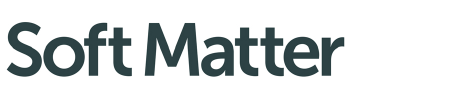}\hfill\raisebox{0pt}[0pt][0pt]{\includegraphics[height=55pt]{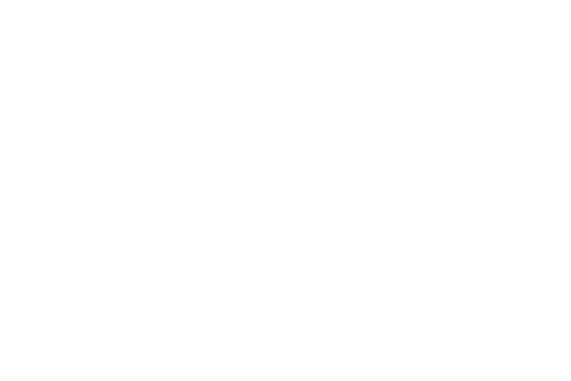}}\\[1ex]
\includegraphics[width=18.5cm]{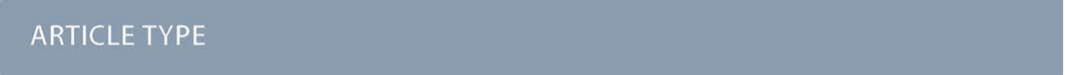}}\par
\vspace{1em}
\sffamily
\begin{tabular}{m{4.5cm} p{13.5cm} }

\includegraphics{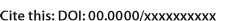} & \noindent\LARGE{\textbf{Dynamics and rupture of doped Motility Induced Phase Separation}$^\dag$}
\footnotetext{\dag~Supplementary Information available: [details of any supplementary information available should be included here]. See DOI: 10.1039/cXsm00000x/} \\%Article title goes here instead of the text "This is the title"
\vspace{0.3cm} & \vspace{0.3cm} \\

 & \noindent\large{Rodrigo Fernández-Quevedo García\textit{$^{a,b}$} Enrique Chacón,\textit{$^{c}$} Pedro Tarazona\textit{$^{d}$} and Chantal Valeriani\textit{$^{a,b}$}} \\%Author names go here instead of "Full name", etc.

\includegraphics{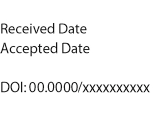} & \noindent\normalsize{Adding a small amount of passive (Brownian) particles to a two-dimensional dense suspension of repulsive Active Brownian Particles does not affect the appearance of a Motility-Induced Phase Separation into a dense and a dilute phase, caused by the persistence of the active particles' direction of motion.  
%the spontaneous condensation of active particles due to the persistence of their direction of motion.   In a purely active suspension, the so-called  Motility-Induced Phase Separation (MIPS) only appears  at relatively high activity and density and leads to the formation of a dense slab in an elongated box under periodic boundary conditions.
Unlike a purely active suspension, the dense slab formed in an elongated system of a passive-active mixture may show, over long periods of time, a stable and well-defined propagation of the interfaces, because of the symmetry breaking caused by the depletion of passive particles on one side of the slab. 
We investigate these dynamical structures via average density profile calculations, revealing an asymmetry between the two interfaces, and enabling a kinetic analysis of the slab movement. 
The apparent movement of the dense slab is not a pure source/sink effect, nor a rigid displacement of all the particles, but a self-sustained combination of both effects. Furthermore, we analyse the specific
fluctuations that produce, cancel and abruptly reverse the slab motion.
%Adding a small amount of passive (Brownian) particles to a two-dimensional suspension of Active Brownian Particles does not alter the spontaneous condensation of active particles due to the persistence of their direction of motion.   
%In a purely active suspension, the so-called  Motility-Induced Phase Separation (MIPS) only appears  at relatively high activity and density and leads to the formation of a dense slab
%in an elongated box under periodic boundary conditions.
%However, differently from  a purely active suspension, the slab formed for a passive-active mixture  is characterised by  a stable and well-defined propagation of the interfaces due to the symmetry breaking caused  by the depletion of passive particles at one side of the slab.
%We thoroughly investigate this structure using 
%average density profiles, revealing  an asymmetry between the two interfaces, and enabling a kinetic analysis of slab movement. 
%This analysis helps to characterize whether the movement of the dense phase is induced by a source/sink effect or the presence of currents within the system itself. Furthermore, significant fluctuations have been observed in the system, capable of either disrupting the slab or abruptly changing its direction of motion due to the nucleation of low-density bubbles within the slab.
} \\%The abstrast goes here instead of the text "The abstract should be..."

\end{tabular}

 \end{@twocolumnfalse} \vspace{0.6cm}
]

%%%END OF TITLE, AUTHORS AND ABSTRACT%%%

%%%FONT SETUP - please do not change any commands within this section
\renewcommand*\rmdefault{bch}\normalfont\upshape
\rmfamily
\section*{}
\vspace{-1cm}

%%%FOOTNOTES%%%

\footnotetext{\textit{$^{a}$~Departamento de Estructura de la Materia, Física Térmica y Electrónica, Universidad Complutense de Madrid, 28040 Madrid, Spain}}

\footnotetext{\textit{$^{b}$ GISC - Grupo Interdisciplinar de Sistemas Complejos 28040 Madrid, Spain}}

\footnotetext{\textit{$^{c}$~ Instituto de Ciencia de Materiales de Madrid (ICMM), Consejo Superior de Investigaciones Científicas (CSIC), Campus de Cantoblanco, 28049 Madrid, Spain}}%

\footnotetext{\textit{$^{d}$~ epartamento de Física Teórica de la Materia Condensada, Condensed Matter Physics Center (IFIMAC), Universidad Autónoma de Madrid, 28049 Madrid, Spain}}%

%Please use \dag to cite the ESI in the main text of the article.
%If you article does not have ESI please remove the the \dag symbol from the title and the footnotetext below.
\footnotetext{\dag~Supplementary Information available: [details of any supplementary information available should be included here]. See DOI: 10.1039/cXsm00000x/}
%additional addresses can be cited as above using the lower-case letters, c, d, e... If all authors are from the same address, no letter is required

%\footnotetext{\ddag~Additional footnotes to the title and authors can be included \textit{e.g.}\ `Present address:' or `These authors contributed equally to this work' as above using the symbols: \ddag, \textsection, and \P. Please place the appropriate symbol next to the author's name and include a \texttt{\textbackslash footnotetext} entry in the the correct place in the list.}

%%%MAIN TEXT%%%%

%%%%%%%%%%%%%%%%%

%\textcolor{blue}{ Active particles continuously dissipate energy to sustain their motion, which makes them to be far out-of-equilibrium. The significance of such systems arises from their complex phenomenology, including emergent behaviors driven by the intrinsic dynamics of out-of-equilibrium systems, and their frequent association with living matter.}
%Suspensions of active particles display interesting collective phenomena, not observed in equilibrium systems.

\section{Introduction}

One of the simplest theoretical models proposed to unravel the features of the collective behaviour of  active particles is the  so-called Active Brownian Particles (ABP) model, consisting of self-propelling  Brownian particles which  gradually change their direction of motion \cite{Cates2013,Gomez2018,Digregorio2018,Caporusso2020}. 
\textcolor{black}{Numerical simulations of suspensions of repulsive ABP have demonstrated that} 
%at a relatively high density and activity,  %numerical simulations of suspensions of %repulsive ABP have been shown  to
\textcolor{black}{ABP spontaneously aggregate due to the persistence of particles' direction of motion, undergoing a  Motility Induced Phase Separation (MIPS)\cite{Stenhammar2014,Bialke2015,Omar2020,Bechinger2016,Redner2013}  into a dense and a dilute phase \cite{MIPScates,Stenhammar2014,Wysocki_2014,Roca_2021,Mason_2023,Digregorio2018,Redner2013,Shi2019,Levis2017,Caprini2019,Solon2018}.} 
\textcolor{black}{These numerical results have been  supported by phase-field calculations \cite{Wysocki_2016,Tjhung2018,Schmidt_2021,Gonnella2015} and experimental results on two-dimensional suspensions of active colloids \cite{Buttinoni2013,Thutupalli2011,Nishiguchi2015,Briand2016}.}

\textcolor{black}{When MIPS takes place in numerical simulations, dealing with an elongated cylindrical box allows characterizing its structural features, such as its interfacial properties.
As suggested by the authors of Ref.\cite{Bialke2015-ko}, considering the swim pressure as a contribution to the total pressure results in  a negative interfacial tension. However, differently from equilibrium, this leads to a long-time stable MIPS.} 
%differently from equilibrium,   in out-of-equilibrium  it   leads to the MIPS long-time stability. 
Patch and coworkers \cite{Patch} have addressed the controversy of a negative surface tension  coexisting with a stable interface by discovering a Marangoni-like effect arising from the presence of sustained tangential currents at the interfaces (on both  dilute and dense phases). 
By means of a continuum description, the authors of Ref.\cite{Lauersdorf_2021} demonstrated that  modelling activity as a spatially varying force allows one to predict consistent pressures and nearly zero surface tension.

\textcolor{black}{
Hermann and coworkers \cite{Hermann2019} implemented a 
a non-equilibrium generalization of the microscopic treatment of the interface,  leading to a positive interfacial tension that directly explained the stability of the interface. 
The approach by  \cite{Hermann2019} produces the
"intrinsic density profile" \cite{chacon2005}. Thus, one could  study capillary wave fluctuations and the wave vector dependence of the interfacial tension. }
Some of us have recently  analysed  the features of  the dense/dilute interfaces  in terms of intrinsic density and force profiles, calculated by means of Capillary Wave Theory \cite{Chacon2022}. 
In our work, we attributed the MIPS stability   to a local rectification of the random active force acting on  particles  at  the dense (inner) side of the MIPS  interface. This caused an external potential  producing a pressure gradient across the interface and lead us to  conclude that  the MIPS mechanical surface tension  could not be described as the surface tension  of equilibrium coexisting phases.  

\textcolor{black}{The stability of  MIPS has been tested against several  features.  On the one side, 
hydrodynamics, together with the particles' shape, has been shown to affect the existence of MIPS. 
MIPS is suppressed when  particles are spherical \cite{Matas-Navarro2014}, while it is enhanced when particles  are elongated \cite{Theers_2018}. 
MIPS is also hindered  in the case  of  a suspension of chemotactic Active Brownian Particles \cite{Hongbo_2023}.}
\textcolor{black}{On the contrary, MIPS is neither  hindered  when particles' equations of motion are characterised by inertia\cite{lowen2020,su2021} nor when particle's motion is on-lattice \cite{barriuso2021,Dittrich2021} rather than off-lattice.}

Adding  passive particles to an ABP suspension,  one could study the physics of a binary mixture. 
By means of  experiments or numerical simulations, mixtures of active and passive particles have been studied by several authors \cite{Dikshit2022,Ai_2020,Wang_2020,Hauke_2020,Marenduzzo_2011,Libchaber_2000,Tu_2001,Patteson2018}. 
\textcolor{black}{When focusing on the behaviour of passive particles,} 
active particles (whether ABP or bacteria) influence not only the structural arrangement of passive particles, but also   their dynamical properties, as demonstrated in 
%suspensions of  Active Brownian Particles and passive colloids 
\cite{Wang_2020,Hauke_2020} and
% bacterial baths and passive colloids 
 \cite{Marenduzzo_2011,Libchaber_2000,Tu_2001,Patteson2018}. 
\textcolor{black}{On the other side, adding a small amount of passive particles to a  suspension of active particles \cite{rogel2020,Wysocki_2016,Cates_2015} is not enough to impede MIPS to take place but can alter its morphology.} 
%It has been shown \cite{Cates_2015,rogel2020,Wysocki_2016} that  %a given amount of 
%passive particles do not hinder  MIPS formation. %However, differently from the purely active system, 
\textcolor{black}{When MIPS appears, the dense phase is  not homogeneous anymore, since  active  particles  are more concentrated %predominantly 
at the boundaries and  passive ones mostly concentrated at its inner part.}
\textcolor{black}{Interestingly, the fluctuations of the MIPS interface are much more pronounced in the binary mixture than in the purely active system\cite{rogel2020}.}
%the  dynamics of the MIPS interface is more violent than in the purely active case, showing  larger fluctuations. 
%Studying the same binary mixture of repulsive Active and passive  Brownian Particles  as in  \cite{Wysocki_2016}, Rogel and coworkers \cite{rogel2020}   demonstrated that MIPS is stable against the presence of passive particles up to very high passive particle concentrations and confirmed that  fluctuations of the MIPS phase are strongly affected by the presence of the passive particles. 
In this respect,  Wysocki and coworkers \cite{Wysocki_2016} studied an active/passive binary mixture in  a two dimension elongated simulation box 
and identified a collective motion of the dense MIPS phase.  
%when preparing the system in a two dimension elongated simulation box. 
This collective motion,  completely absent  in the purely active ABP suspension\cite{Cates_2015},  consisted in well-defined propagating interfaces caused by a flux imbalance of the active and passive particles in the dilute phase.

\textcolor{black}{
In the present work, we investigate the behaviour of a two dimensional binary mixture of  repulsive (WCA)
active ($A$)/passive ($P$) Brownian particles via simulations by means of the   \textit{LAMMPS} Molecular Dynamics package~\cite{LAMMPS}.} 
\textcolor{black}{
The Active Brownian particles, at temperature $T$, with Boltzmann constant $k_B$, translational diffusion coefficient $D_{t}$, and the stochastic uncorrelated white noise $\vec{\xi_{i}}$, have velocities given by
\begin{equation}\label{Eq_movimiento}
    \dot{\vec{r}}_{i}=\dfrac{D_{t}}{k_{B}T}\left(-\sum_{j\neq i}\vec{\nabla} U(r_{ij})+|F_{a}|\vec{n}_{i}\right)+\sqrt{2D_{t}}\vec{\xi}_{i},
\end{equation}
where on top of the interaction  with all  other particles, through 
the pair potential $U(r)$, one should consider a constant self-propelling force, $F_{a}$, whose direction is given by the unitary vector $\vec{n}_{i}=(\cos{\theta_{i}}, \sin{\theta_{i}})$, with the angle $\theta_{i}$ changing independently for each particle, 
\begin{equation}\label{Eq_Evol_direccion_activa}    \dot{\theta_{i}}=\sqrt{2D_{r}}\eta_{i},
\end{equation}
with the rotational diffusion coefficient, $D_{r}$,  and the stochastic uncorrelated white noise $\eta_{i}$.}

\textcolor{black}{
Using the same Eq. (\ref{Eq_movimiento}), the passive particles are simulated by setting $F_{a}=0$. The interaction potential in the first term of the Eq. (\ref{Eq_movimiento}), without distinction between the active or passive character of  particles, is chosen as the repulsive  WCA potential \cite{WCA},
\begin{equation}\label{Eq_WCA}
    U(r)=\left\{
\begin{array}{rcl}
 & 4\epsilon \left(\left(\dfrac{\sigma}{r}\right)^{12}-\left(\dfrac{\sigma}{r}\right)^{6}\right)+\epsilon\hspace{0.5cm}  r<2^{1/6}\sigma \\
 & 0 \hspace{4.15cm}  r>2^{1/6}\sigma
\end{array}
    \right.
\end{equation}
where $r$ is the distance between the centre of two particles and $\epsilon$ is the minimum value of the well of a Lennard-Jones potential. All quantities  presented in this paper are expressed in reduced units, in terms of $\sigma$ and $  \epsilon$.  
%The time step used in all simulations is $\delta t=5\times 10^{-5}$. 
The time step for Brownian dynamics simulations is fixed at $\delta t = 5 \times 10^{-5}$, and simulations typically run for up to $5 \times 10^{3}$ time units, with longer runs ($10^{4}$) performed when higher statistical accuracy is required.}

\textcolor{black}{
In our simulations, we  keep    
%both the temperature $T$ and 
the area $A$  applying periodic boundary conditions in a box with lengths $L_{x}> L_{y}$ (as in \cite{rogel2020}). The periodicity along the short edge ($L_y$)  plays the usual role of avoiding finite size effects; however, on the long edge  ($L_x$)  its role should be regarded as a system with real cylindrical geometry. }
\textcolor{black}{
We set the number of active ($N_{\hbox{\tiny A}}$) and passive ($N_{\hbox{\tiny P}}$) particles constant, so that the total number of particles is $N_{\hbox{\tiny T}} = N_{\hbox{\tiny A}} + N_{\hbox{\tiny P}}$. Unless explicitly stated otherwise, we set $N_{\hbox{\tiny T}} = 8055$ with a box size of $L_x = 200\sigma$ and $L_y = 50\sigma$, where $\sigma$ denotes the particle diameter. 
%In mixtures, $N_{\hbox{\tiny T}}$ is composed of $N_{\hbox{\tiny A}}$ active and $N_{\hbox{\tiny P}}$ passive particles, 
The  passive particles concentrations varies from  $\eta_{\hbox{\tiny P}} = N_{\hbox{\tiny P}}/N_{\hbox{\tiny T}} = 0.4$ down to  the purely active system ($\eta_{\hbox{\tiny P}} = 0$). Throughout the study, the number density is fixed at $\rho = N_{\hbox{\tiny T}}/(L_x L_y) = 0.8055$. Almost most of the results correspond to the standard system size ($L_x = 200\sigma$, $L_y = 50\sigma$, $N_{\hbox{\tiny T}}=8055$). However, when explicitly indicated, we will also present results for other system sizes. 
%All simulations are performed at constant density and run until a steady state is reached.
}

\textcolor{black}{
The activity is controlled via the Peclet number, defined as $P_{e} = 3\nu \tau_{r}/\sigma$, where $\nu = F_{a} D_{t} / k_{B}T$ is the self-propulsion velocity and $\tau_{r} = 1/D_{r}$ is the reorientation time. We set $D_{t} = 1.5$, $D_{r} = 0.6$, $k_{B}T = 1.5$, and $F_{a} = 24$, resulting in $P_{e} = 120$. This choice of density and Peclet number guarantees phase separation (MIPS) for both the pure system ($\eta_P = 0$) and the mixtures studied here, up to $\eta_P = 0.4$ \cite{rogel2020,Cates_2015}. }

\section{Characterization of the dense slab motion}

\begin{figure}[h!]
\centering
\includegraphics[width=\linewidth]{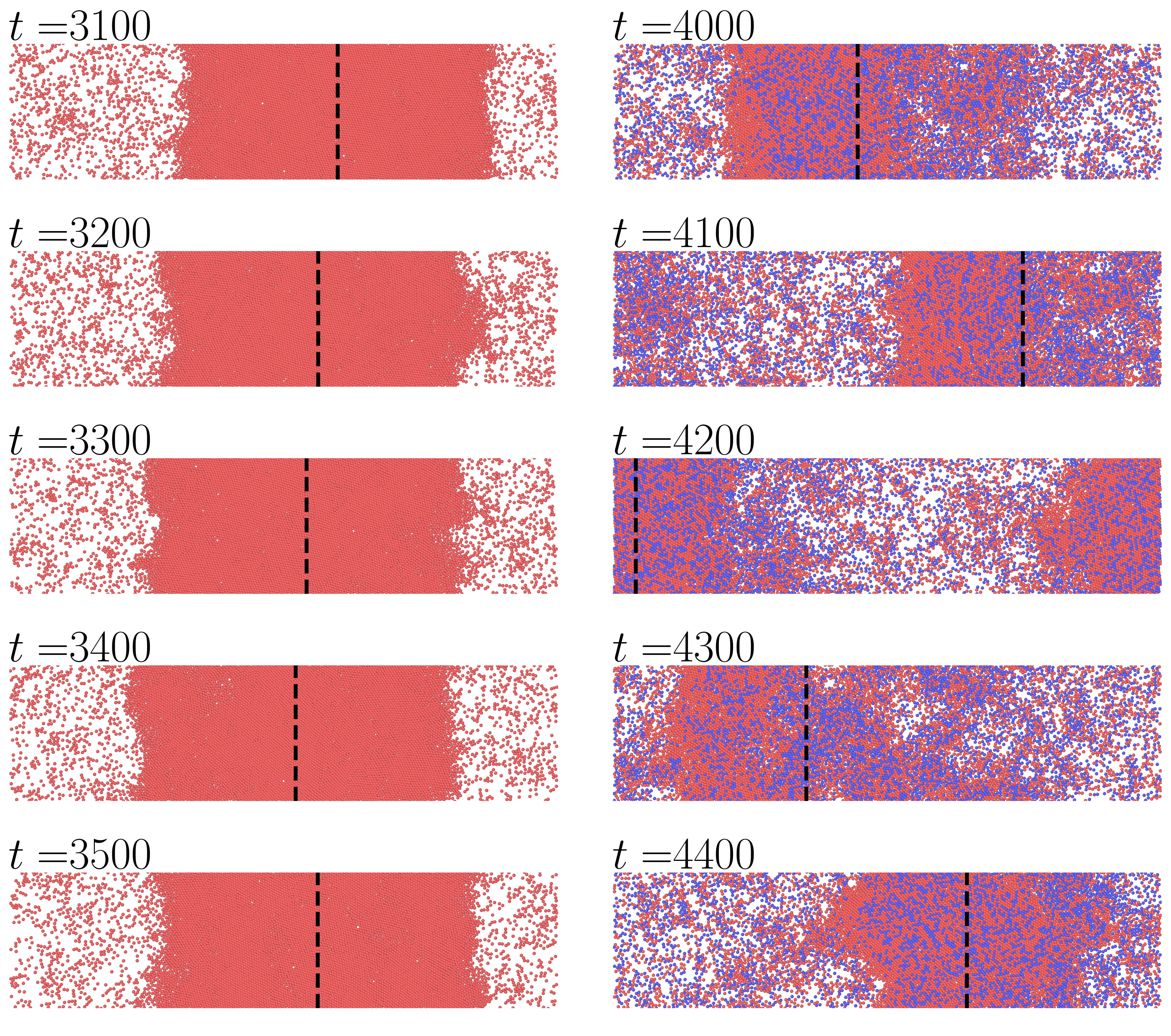}
\caption{Top: MIPS snapshots for $\eta_{P}= 0$ (left-hand side) and $\eta_{P}= 0.4$ (right-hand side) (active particles in red and passive ones in blue). The time difference between the top and bottom snapshots is $\Delta t=100$. \textcolor{black}{(time indicated in each frame) The black dashed lines mark the position of the slab's center of mass}. %as indicated in the legend. 
}
\label{figure1}
\end{figure}

Typical MIPS snapshots, separated by $\Delta t=100$ time intervals
(from top to bottom) are reported in 
Figure \ref{figure1} %-top panels) 
for the  pure $\eta_P=0$ system (left) and the $\eta_P=0.4$ mixture (right), (see also videos$^\dag$ Vid\_SP\_mov\_MIX\_00.mp4 and Vid\_SP\_mov\_MIX\_04.mp4).

\textcolor{black}{The MIPS in the pure system of ABP
(left panels) forms a dense slab (or band) closing itself over the $L_y$ (shorter) period, and filling a fraction $L_{s}/L_{x}$ of the (horizontal, in the Figure) $X$ axis. The less dense phase fills the rest ($(L_x-L_s)\times L_y$) of the simulation box. % fluctuating interfaces. 
The periodic boundary along $X$ implies that the dense band may appear at any position, even apparently split in two pieces, at the opposite ends of the $0\leq x\leq L_x$ box. As usual in phase separations, the thickness of each region would readjust to the values of $N_{\hbox{\tiny T}}$ and the area
$A=L_x \times L_y$, 
%if we change the number of particles and the length $L_x$, 
without change in the coexisting densities.
%The periodic boundaries  may be regarded as a computational trick, to reduce the finite size effects, and we expect that any fixed physical boundaries at $x=0$ and $x=L_x$ would not affect to the coexisting MIPS densities in pure ABP system.}
}

\textcolor{black}{
For the $\eta_P=0.4$ mixture (right column in Figure \ref{figure1}) 
%the periodic boundaries become more relevant.  
the snapshots show also the formation of a dense slab, although thinner 
than at the left column, because in the mixture there is less 
density difference between the  two phases. That could be expected, since 
the active particles are diluted (keeping the same total density) in the mixture with passive particles.
The fluctuations at the interfaces are larger in the mixture than in the pure ABP system, but the most important difference comes in the series of snapshots along each column (see also the videos$^\dag$).}

\textcolor{black}{
In the pure active system, the dense MIPS band has a slow random wandering along the $X$ direction, while in the mixture there is fast translation of the band along the $X$ direction, either to the right (as shown in these snapshots) or to the left (at other times along the simulation).  The same sense of motion is
kept over long time periods and the band often goes across the periodic boundary $x=0$, $L_x$.}

\textcolor{black}{
As shown below, this self-sustained movement relies on
the inhomogeneity of the whole system along the $X$ direction, including a clear asymmetry between the interfacial structures at the two edges of the moving slab. That contrast with the homogeneous phases in the MIPS of pure ABP systems, in which the inhomogeneity is restricted to the two symmetric interfacial regions.}

\textcolor{black}{
The formation of inhomogeneous structures with steady movement is known for systems of Brownian interacting particles kept (by an external force) in steady motion with respect to the bath, so that the balance between the
friction, the external force and the interactions creates inhomogeneous "front" and "wake" 
structures along the direction of motion \cite{Penna}.}

\textcolor{black}{
The use of periodic boundary conditions
along the $X$ axis (perpendicular to the interfaces) becomes then a physical choice (like in a real cylindrical or toroidal geometry), rather than a computational trick, since
the values of $L_x$ and $L_s$ (that is controlled by $N_{\hbox{\tiny T}}$) affect to the whole inhomogeneous  structure.
Taking  $L_x$ as a physical period allows to explore steady states in which the inhomogeneous density distributions move 
(over long time periods) without changing its mesoscopic structure.}

\textcolor{black}{
Our analysis, under these simplified conditions, aims to
quantify the (visually observed) difference between pure and mixed systems. The 
relevance for more complex mesoscopic MIPS structures in mixtures under other geometrical constrains would be discussed later.}

\textcolor{black}{First we need a quantitative mesoscopic description for the motion of these structures, averaging over the rapid changes of the particle positions $x_i$ (for $i=1,N_T$). To that effect} we define (see SM$^\dag$ for discussion on alternative definitions) the instantaneous  position of the dense band $X(t)$ as 
\footnote{Along the simulation 
we chose the branch of the arc-tangent in which $X(t)$ is the closest to its previous position $X(t-\delta t)$, in order to keep the number full $L_x$ rounds.}
\begin{equation}
X(t)=\frac{L_{x}}{2\pi} \arctan{\left(\frac{\sum_{i=1}^{N} \sin\left(\frac{2\pi}{L_{x}} x_{i}(t)\right)}{\sum_{i=1}^{N} \cos\left(\frac{2\pi}{L_{x}} x_{i}(t)\right)}\right)},
 %+\frac{L_{x}}{2}
\label{Eq_Mass_Centre_Fourier}
\end{equation}
\textcolor{black}{that corresponds to get real positive values for the lowest ($q=2\pi/L_x$) Fourier component of the particle position $\sum_{i=1}^{N_{\hbox{\tiny T}}} e^{i q (x_i(t)-X(t))}$; so that $X(t)$ is located at the center of the dense slab, independently of the position of the slab in the $0\leq x\leq L_x$ interval.}
\begin{figure}[h!]
\centering
\includegraphics[width=\linewidth]{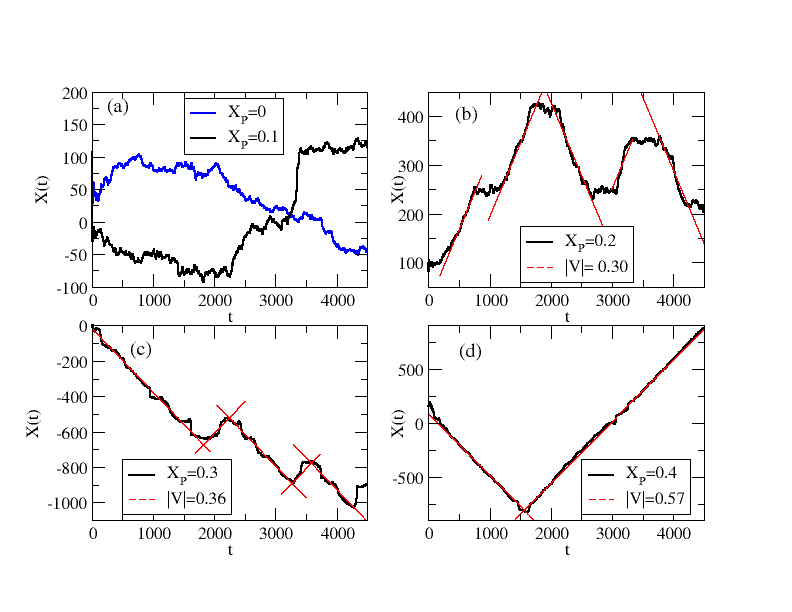}
\caption{\textcolor{black}{Slab position over time, $X(t)$, for various $\eta_{P}$ values (see legend). Red dashed lines indicate periods of linear displacement, with slopes (velocities $|V|$) also shown in the legend.}}
\label{figure1bis}
\end{figure}

The results for $X(t)$, reported in Figure \ref{figure1bis}),
\textcolor{black}{are then used to calculate the velocity of the slab as a mesoscopic structure
%$V(t)$ is obtained from  the 
%change of $X(t)$ between time steps, $\Delta t=0.05=10^3\delta t$, used to analyse the molecular positions along the simulation,
\begin{equation}\label{Eq_SP_velocity}
    V(t)=\frac{dX(t)}{dt}\approx\frac{X(t+\Delta t)-X(t)}{\Delta t},
\end{equation}
with time step $\Delta t=0.05=10^3\delta t$.
}

\textcolor{black}{
Then $V(t)$ is averaged as $\langle V\rangle$, over $250\Delta t=12.5$ time intervals, and
used to get its autocorrelation$^\dag$, $\langle (V(t_o)-\langle V\rangle)(V(t_o+t)-\langle V\rangle)\rangle\approx \Delta V^2 exp(-t/\tau_p)$, to characterize 
rapid fluctuations with persistence time $\tau_{p}\approx 1.8\pm 0.2$, 
that we find to be similar in pure and mixed systems.}
%($0\leq  \eta_P\leq 0.4$). 

\textcolor{black}{
For the unbiased random walk ($\eta_P=0$) there is no longer time scale than $\tau_p$, so that
the mean square amplitude of the $V(t)$ fluctuations $\Delta V^2\equiv \langle(V(t)-\langle V\rangle)^2\rangle$ should produce a random walk of $X(t)$ with diffusion constant 
$D=\Delta V^2 \tau_p \approx 0.3\pm 0.05$; 
consistent with the observed changes $\Delta X(t) \sim \pm 50$ over the full simulation 
run $t_{max}=4400$. Such slow drift (that should be size dependent) is typical in any phase separation and it is usually dealt with (small) rigid shifts of all  particle positions to 
keep the dense slab at an approximately fixed position in the periodic cell.}

\textcolor{black}{In contrast, mixtures with
$\eta_p\geq0.2$ (Figure \ref{figure1bis}, panels b-d) show long Steady Motion Periods (SMP), marked as red dashed lines, with a
nearly constant (either positive or negative) slope $X(t)-X(t_o) \approx \pm V_{\hbox{\tiny SMP}}(\eta_P) \  (t-t_o)$, 
with (absolute) values of the slope that increase from $V_{\hbox{\tiny SMP}}(0.2)\approx 0.30$ to $V_{\hbox{\tiny SMP}}(0.4)\approx 0.57$.
The mean square fluctuations $\Delta V^2=\langle (V(t)-\langle V(t)\rangle)^2\rangle\approx 0.3$ (over the same $12.5$ time intervals) are just slightly larger than for $\eta_P=0$.}

\textcolor{black}{
The amplitude $\sqrt{\Delta V^2}\approx 0.5$ of these fast $\tau_p\approx 1.8$ fluctuations is of the same order of magnitude as the mean value $V_{\hbox{\tiny SMP}}$; 
however the SMP are clearly identified in $X(t)$ (see Figure \ref{figure1bis}) and also in the 
histograms for $V(t)$$^\dag$) because a much longer time scale emerges, with a
typical SMP persistent time $\tau_{\hbox{\tiny SMP}}$ that we can only estimate 
to be $\tau_{\hbox{\tiny SMP}}\sim 500$ for $\eta_P=0.2$,  $\tau_{\hbox{\tiny SMP}}\sim 1000$ for $\eta_P=0.3$ and larger
for $\eta_P=0.4$}

\textcolor{black}{
The SMPs are interrupted by %sharp changes in $X(t)$ that indicate a 
large fluctuations in the structure of the MIPS slab, reflected in 
sharp changes of the nominal position $X(t)$ defined in Eq.(\ref{Eq_Mass_Centre_Fourier}). Then, there
may be a rest period, in which $X(t)$ fluctuates without forth-back bias, before a new sudden change in $X(t)$ leads to start a new SMP, that may be in the opposite
direction (with a U-turn of the slab) or in the same direction as the previous SMP.
We may estimate that for $\eta_P=0.4$ there is a typical time $\tau_o \sim 100$ (just order of magnitude) for the rest period between two SMP.}
\textcolor{black}{U-turns occur at each $\eta_p$ during periods of steady motion. As $\eta_p$ increases, the time required for a U-turn to develop becomes shorter.  Thus, at $\eta_p=0.4$, the U-turn is very abrupt. In this respect, given sufficient simulation time, a U-turn d also occurs at $\eta_p=0.3$, although less sharply than at $\eta_p=0.4$, but still more sharply than at $\eta_p= 0.2$.}

\textcolor{black}{
Therefore, for that high concentration of passive particles $\tau_p\ll\tau_o\ll\tau_{\hbox{\tiny SMP}}$ gives a good separation of time scales for our analysis of the SMPs.  For lower $\eta_P$,  $\tau_o$ increases while $\tau_{\hbox{\tiny SMP}}$ decreases, so that for $\eta_P=0.2$ we may estimate
$\tau_o\sim\tau_{\hbox{\tiny SMP}}\sim 500$.}

\textcolor{black}{
Notice that we have only very crude estimations for $\tau_{\hbox{\tiny SMP}}(\eta_P)$ and $\tau_o(\eta_P)$, 
since a good statistics would require extremely long simulation runs. 
As in any mesoscopic
characterization there are uncertainties; a red line in Figure \ref{figure1} may 
go through a large fluctuation, and we could take it as a single SMP or as two 
successive SMP that keep the same direction of motion, after a very short rest time.}

\textcolor{black}{
Nevertheless, we find robust indications for the 
emergence of the time scales $\tau_{\hbox{\tiny SMP}}>\tau_o\gg \tau_p$, leading
to superdiffusive behaviour of the dense MIPS clusters, as a peculiarity of active-passive mixtures and the subject of our study. In the remaining of our work, we will focus on the $\eta_P=0.4$ case
for a detailed analysis, because it provides the best statistics to characterize the 
structures that produce the SMP.}

\textcolor{black}{
However, similar behaviour is clearly observed for $\eta_p=0.3$ and $0.2$. For $\eta_P=0.1$ we observe a sudden changes in $X(t)$, that do not appear in the pure ($\eta_P=0$) MISP slabs, but the time scale separation between SMP and rest periods becomes uncertain.}

\section{Analysis of the density and current profiles}

The instantaneous density profiles 
$\hat{\rho}_\alpha(x,t)=\sum_{i=1}^{N_\alpha} \delta(x-x_i(t))/L_y$, for
$\alpha=A,P$ particles, should be averaged over a time interval
$t\in[t_o,t_o+\Delta t]$ to get 
smooth mean density profiles
$\rho_\alpha(x)=\langle \hat\rho_\alpha(x,t)\rangle_{\Delta t}$. However, 
%for the $\eta_p=0.4$ mixture 
%this is frustrated by
the fast movement of the band in active-passive mixtures %, which
leads to  
flat $\rho_\alpha(x)$, unless the time average is chosen to be very narrow, with
the consequence of too noisy
$\rho_\alpha(x)$.
%: the consequence of this being very noisy results.  
This problem can be overcome by 
evaluating  the shifted density profiles \cite{Stenhammar2014,Roca_2021}
\begin{equation}
\tilde\rho_\alpha(x)=\frac{1}{L_y}\left\langle \sum_{i=1}^{N_\alpha} \delta(x+X(t)-x_i(t))\right\rangle_{\Delta t}.
\label{Eq_mean_density_profile}
\end{equation}

\begin{figure}[h!]
%\begin{figure}[b]
\includegraphics[width=0.79\linewidth]{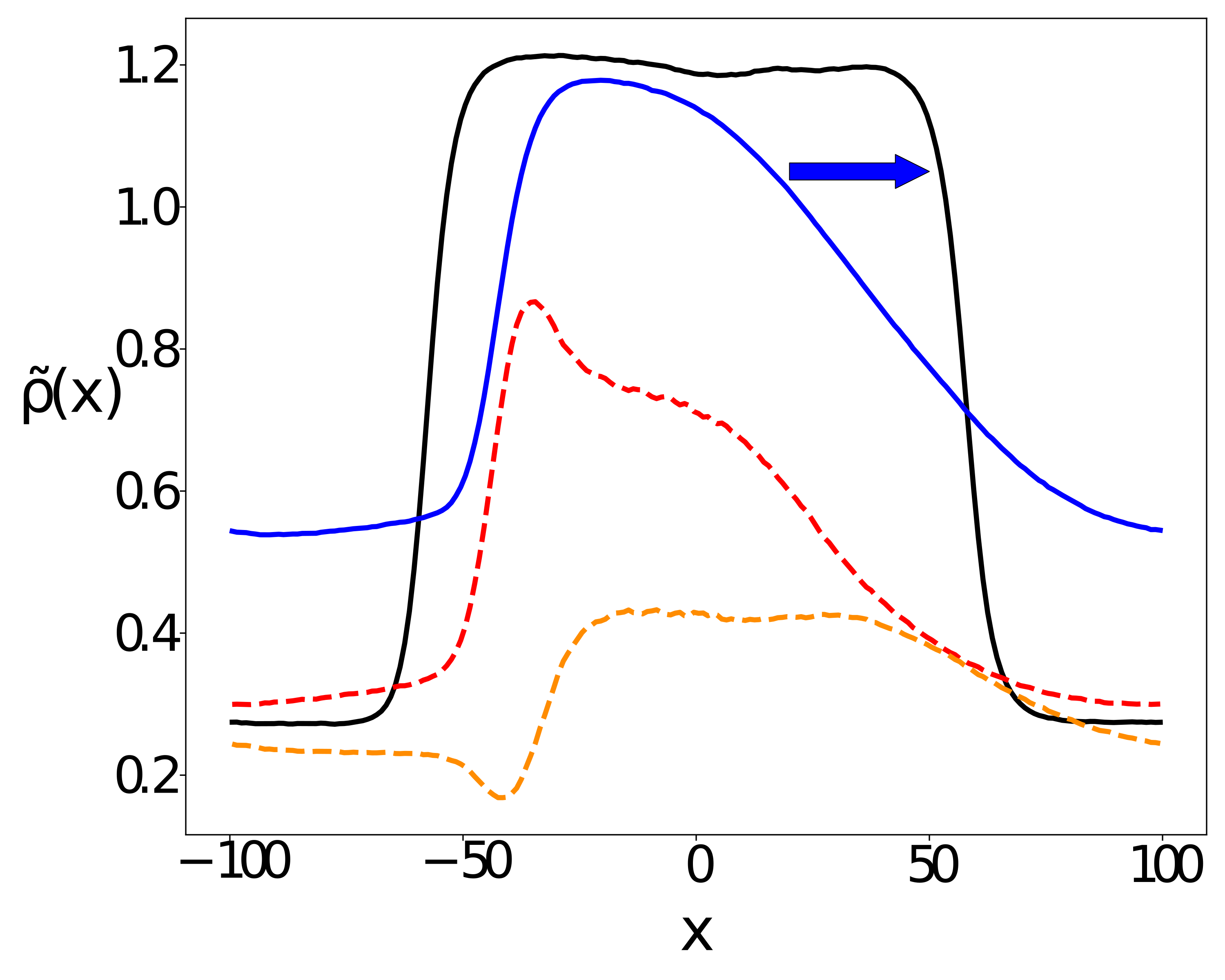}% Here is how to import EPS art
\caption{ Full lines: Mean total
density profile $\tilde\rho(x)$:
%over $\Delta t=4000$,
for  $\eta_{P}=0$ (black) and 0.4 (blue). For the latter, dashed lines: $\tilde{\rho}_A(x)$ (red) and $\tilde{\rho}_P(x)$ (orange)
%the mean density profile for  passive  (dashed orange) and  active particles (dashed red) are plotted, averaged over a SMP in the direction of the blue arrow.
Blue arrow: the direction of motion in the SMP used for the time average.
}
\label{figure2}
\end{figure}

\textcolor{black}{
%We focus first on the $\eta_P=0.4$ system, where we use t
The emergent time scale separation allows to take $\tau_{\hbox{\tiny SPM}}>\Delta t\gg\tau_p$ 
%.For the $\eta_p=0.4$ mixture, we use $\tau_{\hbox{\tiny SMP}} \gg \Delta t \gg \tau_p$ 
with very good average over the rapid fluctuations, but still within a single SMP. 
Thus, for $\eta_p=0.4$, over a period with $\langle V(t)\rangle
\approx +0.57$ we obtain
the strongly asymmetric  $\tilde\rho_A(x)$ and $\tilde\rho_P(x)$  profiles shown in Figure \ref{figure2}.   }

At the back  of the moving band, $\tilde\rho_A(x)$ (dashed red) presents a peak and  $\tilde\rho_P(x)$ (dashed orange) drops \textcolor{black}{to a narrow minimum}. 
%On the contrary, 
In contrast, at the front,  
$\tilde\rho_A(x)$ and 
$\tilde\rho_P(x)$ have similar smoother decays.    
For comparison, the pure ABP system (black line)
presents a fairly symmetric profile and larger MIPS density difference.

\begin{figure}[h!]
%\begin{figure}[b]
\includegraphics[width=1.\linewidth]{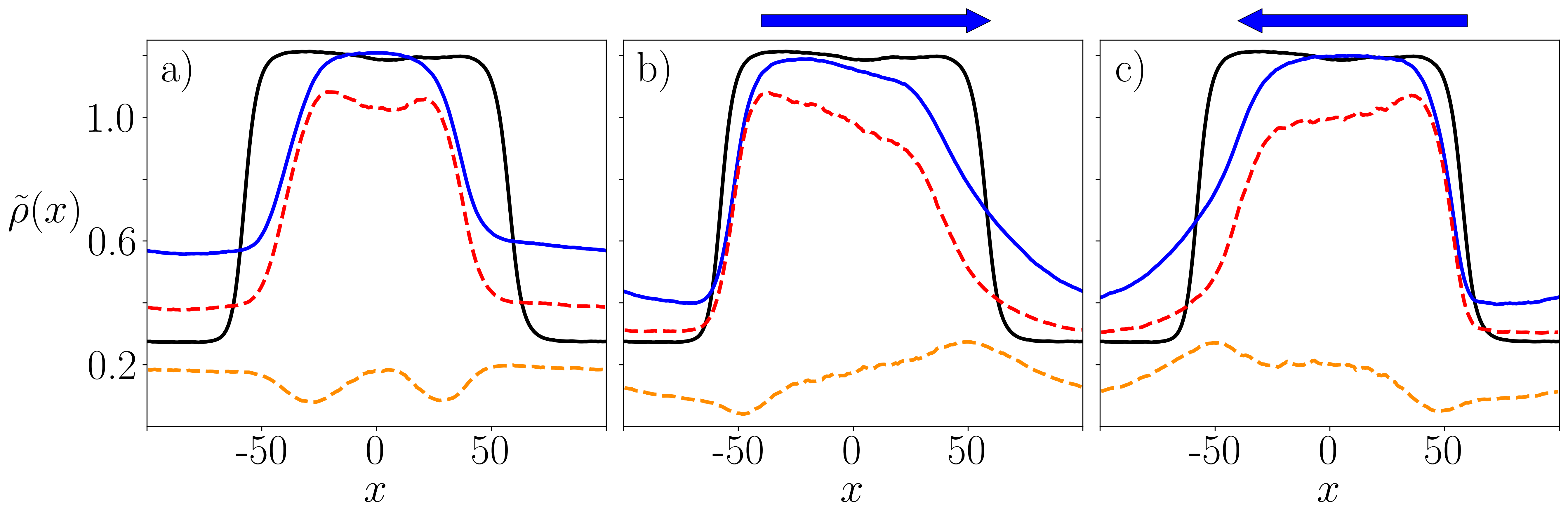}% Here is how to import EPS art
\caption{ Full lines: Mean total
density profiles (Eq.\ref{Eq_mean_density_profile}) for $\eta_{P}=0.2$: Blue line total density,
red-dashed line $\tilde{\rho}_A(x)$ and orange-dashed line
$\tilde{\rho}_P(x)$. For comparison, the black liners gives the density profile in a pure ($\eta_P=0)$ ABP system with the same $\rho_{\hbox{\tiny T}}$ . In panel (a) the average is taken over 
the rest periods ($\langle V(t)\rangle\approx 0$ in panel (b) of Figure \ref{figure1bis}); in panels (b) and (c) the profiles
are averaged over the SMP (red-dashed lines in Figure \ref{figure1bis}) when the slab moves in the direction of the top blue arrow.
}
\label{figure2bis}
\end{figure}

\textcolor{black}{
The same qualitative features are found at lower
concentration of passive particles, as shown in
Figure \ref{figure2bis}, for $\eta_P=0.2$.}

\textcolor{black}{
In panels (b) and (c) Eq.(\ref{Eq_mean_density_profile}) is averaged over the SMP with $\langle V(t)\rangle\approx\pm 0.30$ respectively (along the periods marked as red-dashed lines in Figure \ref{figure1bis}(b)). The asymmetry is
($\tilde\rho_\alpha(x)\rightarrow \tilde\rho_\alpha(-x)$) in SMP with the opposite velocity. The rest periods ($\langle V(t)\rangle\approx 0$) are now long enough to take good  averages, that give the symmetric density profiles in
Figure \ref{figure2bis}(a).}
\textcolor{black}{
Notice that the steady motion 
of the band across the system seems to reinforce (rather than to weaken) the MIPS in the mixture, since the density contrast
between the maximun density (at $x\approx 0$) 
and the minimum density (at $x\approx \pm L_x/2$) is 
higher in the SMP asymmetric profiles, than in the symmetric
profiles during the rest periods.}

%\textcolor{black}{Figure \ref{figure2bis}

\begin{figure}[h!]
\includegraphics[width=1.0\linewidth]{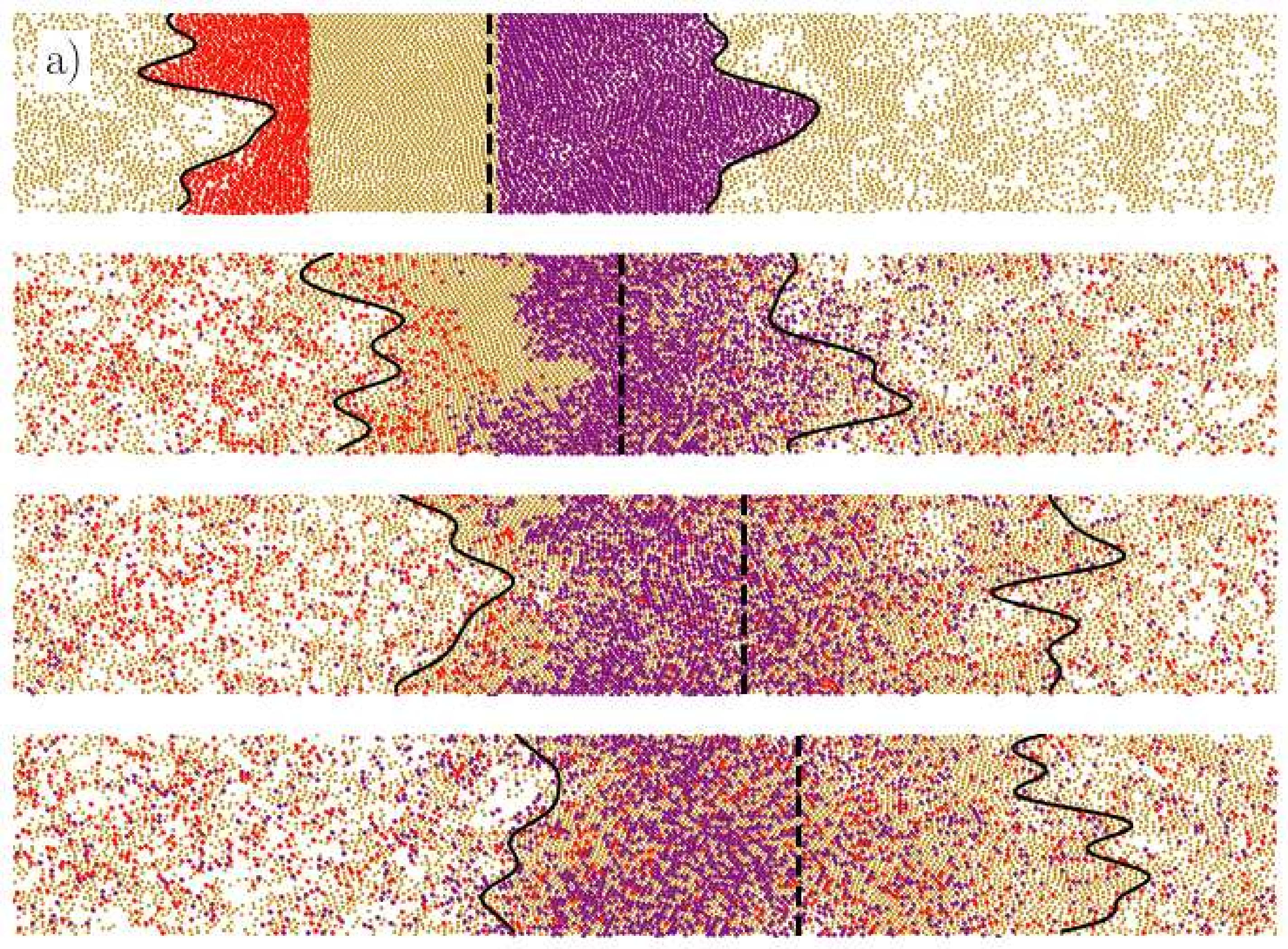}% Here is how to import EPS art
\caption{
Snapshots  showing the slab's movement as a function of  time (from top to bottom). Slab 
boundaries \cite{Chacon2022} underlined by a  black line as well as the mass centre position, $X(t)$ (vertical dashed line). Top snapshot:   slab particles on each side are 
coloured in red (source interface) and purple  (sink interface). Their colour is maintained in the all  snapshots,    showing how particles change position with time.}
\label{figure3}
\end{figure}

Wysocki et al. \cite{Wysocki_2016} proposed a source-sink mechanism to explain the band displacement, 
where the receding interface acts as a source, evaporating particles into the dilute phase, 
while the advancing front, acting as a sink, captures particles into the dense phase. 
A representation of this source/sink effect is reported in Figure \ref{figure3} for %the mixture at 
$\eta_P=0.4$, 
with four snapshots, each with 
%(time  growing from top to bottom), 
$X(t)$ as a 
vertical dashed line and the MIPS boundaries as wavy black lines.

In the top snapshot, particles  
are coloured in red if close to the source (back) side  and in purple if close to the sink (front) one.  
From top to bottom, the red particles rapidly escape from 
the back interface and are added at the front interface. 
The purple particles remain mostly stationary as the slab moves through them, diffusing slowly within the dense region.

For a quantitative analysis,  we use Smoluchowski's equation 
for 
%Brownian particles, with 
the instantaneous distributions of density 
$\rho(\vec{r},t)=\langle \delta(\vec{r}-\vec{r}_i(t)\rangle_n$ and current
$\vec{j}(\vec{r},t)=\langle \delta(\vec{r}-\vec{r}_i(t) \ \vec{v}_i(t)\rangle_n$, 
smoothed by the average $\langle...\rangle_n$ over many realizations of 
the Brownian noise acting on the particles. 
Along a SMP the velocity of the MIPS band has
rapid ($\sim \tau_p$) fluctuations around a mean value $\langle V\rangle$.
 We assume that 
the self-averaging of $\tilde{\rho}(x)$ gives $\rho(\vec{r},t)=\rho(x,t)\approx \tilde\rho(\tilde{x})$, with 
$\tilde{x}=x-\langle V\rangle t$, and
the continuity equation 
for the steady state movement of the band becomes \cite{Penna} 
$\partial_t\rho_\alpha(x,t)=-\partial_x j_\alpha(x,t)=-\langle V\rangle \partial_{\tilde{x}} \tilde{\rho}(\tilde{x})$, 
that may be integrated to the current densities 
%for each species
%, along the $x$ direction and in terms of the variable $\tilde{x}$, takes the form
\begin{equation}\label{Eq_Corriente}    % Ecuacion 3
    j_{\alpha}(\tilde{x}) = \langle V \rangle \tilde\rho_{\alpha}(\tilde{x}) - \Delta_{\alpha}
\end{equation}
where the integration constant, $\Delta_\alpha$,  is 
the counter-current 
%$\Delta_\alpha$ 
%corresponds to the counter-current of each species, that
%gives the difference between 
which makes $j_{\alpha}(\tilde{x})$ to be lower than  
the apparent current carried by the moving slab.

In our simulations, we  get $\Delta_{\hbox{\tiny A}}$ and $\Delta_{\hbox{\tiny P}}$ from $\langle V\rangle$
and the mean velocity \textcolor{black}{of the particles, for each species, 
\begin{equation}\label{Eq_Velocidad_CM_especies}   % Ecuacion 4
    \langle v_{\alpha}\rangle\equiv \dfrac{1}{N_{\alpha}}\biggr\langle \sum_{j=1}^{N_{\alpha}}v_{i} \biggr\rangle=
    \frac{\int_0^{L_x}dx \ j_{\alpha}(x,t)}{\int_0^{L_x}dx \ \rho_{\alpha}(x,t)}
    =\langle V \rangle -\dfrac{\Delta_{\alpha}}{\langle \rho_{\alpha}\rangle }
\end{equation}
}
where $\langle\rho_\alpha\rangle=N_\alpha/(L_xL_y)$ are the fixed mean densities.
%in our simulation box. 
%Considering a  binary mixture with
For $\eta_{\hbox{\tiny P}}=0.4$ and $\langle V\rangle \approx 0.57$ (as in Fig.\ref{figure2}) we get $\langle v_{\hbox{\tiny A}}\rangle=-0.03$ and $\langle v_{\hbox{\tiny P}}\rangle=0.05$, which correspond to counter currents,  $\Delta_{\hbox{\tiny P}}\approx 0.17$ and $\Delta_{\hbox{\tiny A}}\approx 0.29$.  

\section{Brownian dynamics analysis for the SMP}

\textcolor{black}{In the Smoluchowski equation for Brownian particles (with mobility $\Gamma$, temperature $T$ and translational diffusion constant $D_t=kT \ \Gamma$) the 
(noise averaged) currents for each species are 
%determined by the density distributions of force $f_{\alpha}(x,t)$
%plus the diffusion term with the gradient of the density and 
%the translational diffusion constant $D_t=kT \Gamma=kT$,
\begin{equation}\label{Eq_Corriente_Smolu}
    j_{\alpha}(x,t)= \Gamma \left( \sum_{\beta=A,P} f_{\alpha\beta}(x,t)- kT \partial_{\partial x} \rho_{\alpha}(x,t)+ \delta_{\alpha A} \ f^{a}(x,t)\right)
\end{equation}
with the interaction force densities $f_{\alpha\beta}(x,t)$ that particles of species $\alpha$ receive from particles of species $\beta$, the thermal diffusion (both for active and passive particles), and the active force density $f^{a}(x,t)$ (with $\delta_{AA}=1$ and $\delta_{BA}=0$),  which 
reflects the local rectification of the (independently random but slowly varying) active force acting on each particle.}

\begin{figure}[h!]
\includegraphics[width=1\linewidth]{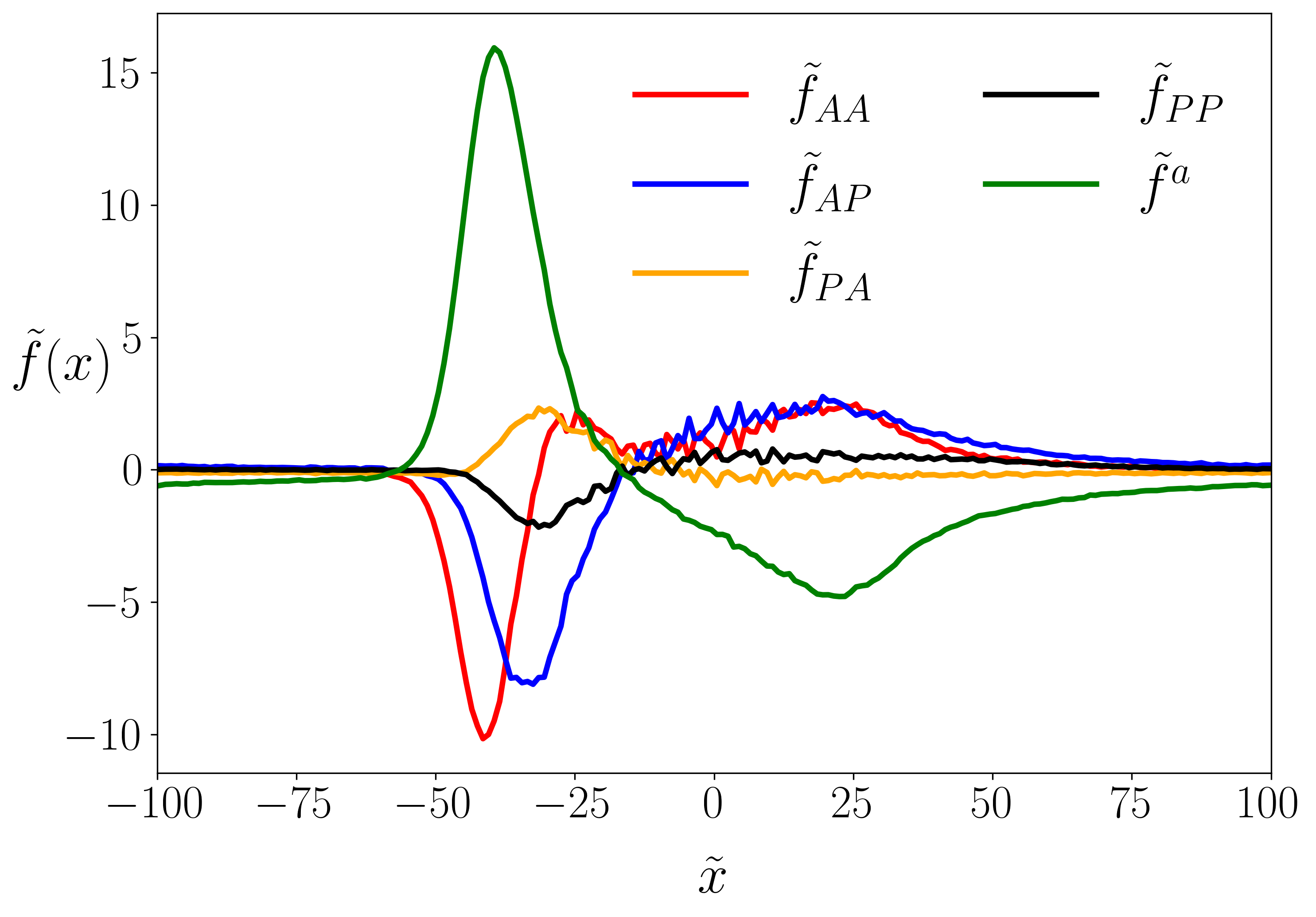}
\caption{Mean force profiles for the system with $\eta_{P}=0.4$ moving towards the positive X-axis: the interaction between active-active, $\tilde{f}_{AA}$ (red), and passive-passive, $\tilde{f}_{PP}$ (black), active-passive, $\tilde{f}_{AP}$ (blue), and the opposite $\tilde{f}_{PA}$ (orange); and the active force density profile, $\tilde{f}^{a}$ (green),
}
\label{figure4bis}
\end{figure}

\textcolor{black}{All these force densities may be evaluated in our simulations, and averaged along a SMP in terms of the relative position $x=x_i(t)-X(t)$ of each particle 
from the mesoscopic position of the slab, as done in Eq.(\ref{Eq_mean_density_profile}) for the density profile $\tilde{\rho}_\alpha(x)$. The results in
Figure \ref{figure4bis} correspond to the same SMP, for $\eta_P=0.4$, as the density profiles
in Figure \ref{figure2}.
}

\textcolor{black}{The assumption $\rho(x,t)\approx \tilde{\rho}(\tilde{x})$, with $\tilde{x}=x-\langle V\rangle t$, that lead to Eq.(\ref{Eq_Corriente}) may now be used
for each component of the force densities
$f_\alpha(x,t)\approx \tilde{f}_\alpha(\tilde{x})$, to take the time averages along a SMP with 
mean slab velocity $\langle V\rangle$ as (self-averaging) representation of the Brownian noise average in
Smoluchowski equation (\ref{Eq_Corriente_Smolu}), to get
\begin{equation}\label{Eq_Corrientes_propto_forces}
    j_{\alpha}(\hat{x}) = \Gamma \left(\sum_{\beta=A,P}f_{\alpha,\beta}(\hat{x})-
    kT \partial_{\tilde{x}}  \tilde\rho_{\alpha}(\hat{x})+
    \delta_{\alpha A} \ f^{a}(\hat{x})\right).
\end{equation}}

\begin{figure}[h!]
\includegraphics[width=1\linewidth]{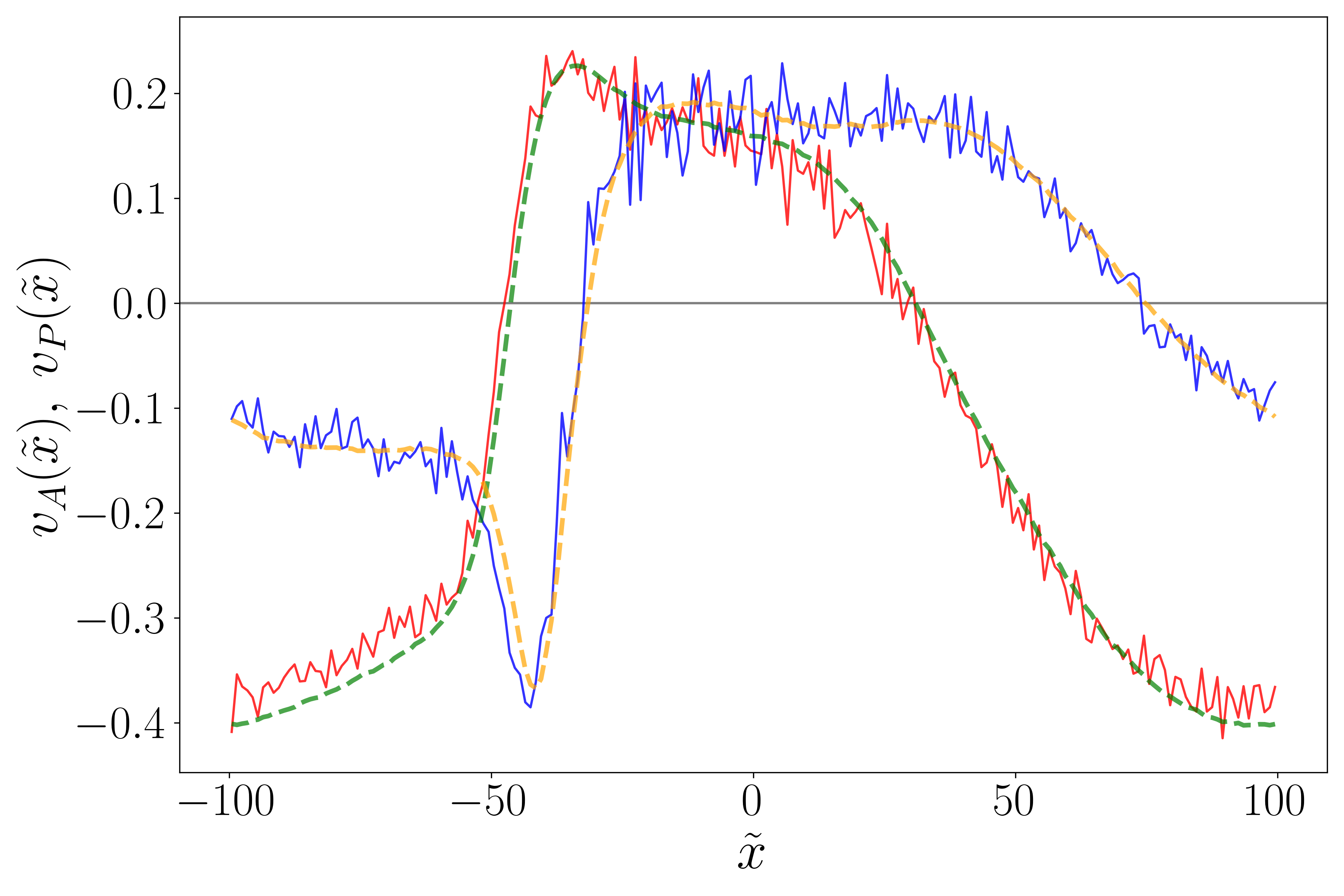}
\caption
{Particles velocity profiles for $\eta_P=0.4$ calculated via Eq. \ref{Eq_Corriente}. Using the currents from the forces (noisy solid lines) and the density profiles (dashed lines). Both the profiles of active (red and green lines) and passive (blue and orange lines) particles are included. The horizontal line represents $v_A(\tilde{x})=v_P(\tilde{x})=0$.}
%Particles velocity profile for $\eta_{p}=0.4$: active particles (red) and passive particles (blue) are represented.}
\label{figure4}
\end{figure}

\textcolor{black}{
Figure \ref{figure4} compares (again for the same $\eta_P=0.4$ SMP analyzed in Figure \ref{figure2}) 
the mean velocities, $v_\alpha(\tilde{x})\equiv j_\alpha(\tilde{x})/\rho_\alpha(\tilde{x})$, obtained via Eq.(\ref{Eq_Corriente}) 
(i.e. from the density profile, the global mean velocities $\langle v_\alpha\rangle$ and 
the slab $\langle V\rangle$) and via Eq.(\ref{Eq_Corrientes_propto_forces})
with the force densities in Figure \ref{figure4bis}.}

\textcolor{black}{
The fairly good agreement
between the kinetic and the dynamical results
(see also  in SM$^\dag$ the direct comparison for $j_\alpha(\tilde{x})$) gives
a strong support to our assumption $j(x,t)\approx\tilde{j}(\tilde{x})$,
i.e. to the self-averaging of the Brownian thermal noise as time averages 
over SMP. Therefore we may proceed with the following dynamical interpretation 
for the structural origin of the SMP based on (\ref{Eq_Corrientes_propto_forces}).
}

\textcolor{black}{
Equating (\ref{Eq_Corriente}) and (\ref{Eq_Corrientes_propto_forces}) gives
\begin{equation}\label{Eq_def_contracorrientes_2}
    \Delta_{\alpha}= \langle V\rangle \rho_{\alpha}(\tilde{x})-
    \Gamma \left( \sum_{\beta=A,P} f_{\alpha,\beta} (\tilde{x})-kT \partial_{x}\rho_{\alpha}(\tilde{x})+f^{a}(\tilde{x})\right).
\end{equation}
The constant value of the right hand side, independent of $\tilde{x}$, is a strong
link between the results for the density and force profiles when the system is in 
a SMP. Moreover, taking the average of (\ref{Eq_def_contracorrientes_2}) over the
total length $0\leq \tilde{x}\leq L_x$, with periodic boundaries, the contributions 
from the Brownian diffusion and the active force (both with independent random direction on each particle) are null.}

\textcolor{black}{
The same applies 
to the repulsion between particles of the 
same species  $f_{\hbox{\tiny AA}}(\tilde{x})$ and $f_{\hbox{\tiny PP}}(\tilde{x})$, 
that have null integrals. 
The qualitative difference of a mixture with respect to a pure ABP system is that
the integrals of $f_{\hbox{\tiny AP}}(\tilde{x})$ and $f_{\hbox{\tiny PA}}(\tilde{x})$,
\begin{equation}
    \frac{1}{L_x}\int\limits_{0}^{L_{x}}dx f_{\hbox{\tiny PA}}(t)=-\frac{1}{L_x}\int\limits_{0}^{L_{x}}dx f_{\hbox{\tiny AP}}(t)
    \equiv \frac{F_{\hbox{\tiny PA}}}{A}=-\frac{F_{\hbox{\tiny AP}}}{A},
\end{equation}
may have non null (but opposite) values
reflecting that, if the density profiles break the left-right symmetry, the active particles may globally 
push (and be pushed back by) the passive ones. 
The static (on average) centre of mass for the whole system, that emanates from the 
Brownian dynamics and the mean average of the active force on each particle, imposes that
$N_{\hbox{\tiny A}} \langle v_{\hbox{\tiny A}}\rangle+ N_{\hbox{\tiny P}} \langle v_{\hbox{\tiny P}}\rangle=0$.}

\textcolor{black}{
As observed in our simulations, the global movement of the two species has
to be in opposite directions, and it is fully determined by their mutual global force $F_{\hbox{\tiny PA}}=-F_{\hbox{\tiny AP}}$. 
%The counter currents of the two species are linked by the restriction
%$\Delta_{\hbox{\tiny A}}+\Delta_{\hbox{\tiny P}}=\langle V\rangle \rho{\hbox{\tiny T}}$, with the total density. 
%
%with the area $A=L_xL_y$ and the global densities
%$\langle \rho_{\alpha} \rangle= N_{\alpha}/A$, this analysis of the Brownian dynamics
The counter currents in Eq.(\ref{Eq_Corriente}) may be obtained in terms of the total forces between active and passive particles,
\begin{equation}\label{Eq_def_contracorrientes_3}
    \Delta_{\alpha}= \langle V(t)\rangle \langle\rho_{\alpha}\rangle-\Gamma \dfrac{F_{\hbox{\tiny PA}}}{A},
    %=(V-\langle v_{\alpha}\rangle)\langle\rho_{\alpha}\rangle
\end{equation}
which amounts to the dynamical balance for the global mean velocity of each species
\begin{equation}\label{Eq_Velocidad_centro_masas_pasivas}
    \langle v_{P}\rangle=\Gamma \frac{F_{\hbox{\tiny PA}}}{N_{\hbox{\tiny P}}}
    \hbox{ \  and  \  }
      \langle v_{A}\rangle=-\Gamma \frac{F_{\hbox{\tiny PA}}}{N_{\hbox{\tiny A}}}
\end{equation}
}
and the static (on average) centre of mass for the whole system,
$N_{\hbox{\tiny A}} \langle v_{\hbox{\tiny A}}\rangle+ N_{\hbox{\tiny P}} \langle v_{\hbox{\tiny P}}\rangle=0$. The global movement of the two species has
to be in opposite directions, and it is fully determined by their mutual global force $F_{\hbox{\tiny PA}}=-F_{\hbox{\tiny AP}}$. 
The counter currents of the two species are linked by the restriction
$\Delta_{\hbox{\tiny A}}+\Delta_{\hbox{\tiny P}}=\langle V\rangle \rho{\hbox{\tiny T}}$, being $\rho{\hbox{\tiny T}}$ the total density. 

\textcolor{black}{The velocity profiles in Fig. \ref{figure4} } show that within the dense slab 
($-25\leq \tilde{x}\leq 15$) the active and passive particles move together, %with similar mean velocities, 
%that are approximately one third of the apparent velocity 
$v_{\hbox{\tiny A}}(\tilde{x})\approx v_{\hbox{\tiny P}}(\tilde{x})\approx \langle V(t)\rangle/3$.
%, i.e. about one third of $\langle V(t)\rangle$. 
Therefore, the MIPS kinetics  in active-passive mixtures is neither a pure source/sink mechanism (in which the particles of the dense slab would remain at rest), nor a real drift of the particles at the full velocity
$V(t)$, as if $\Delta_{\hbox{\tiny A}}=\Delta_{\hbox{\tiny P}}=0$. Drift and source/sink act in parallel,
self-maintained by the asymmetry of $\rho_\alpha(\tilde{x})$, that unbalance the local rectifications of the active forces at the front and back edges of the slab.  With our 
$\rho(x,t)\approx \tilde\rho(x-\langle V\rangle t)$ hypothesis, if $v_A(\tilde{x})=v_P(\tilde{x})$ at a point $\tilde{x}$, their value is
%are set by the total density at that point and its mean value $\rho_{\hbox{\tiny T}}=N/(L_xL_y)$,
%$v_A(\tilde{x})=v_P(\tilde{x})=(1-\rho/(\rho_A(\tilde{x})+\rho_P(\tilde{x}))  \langle V\rangle$.
\begin{equation}
v_{\hbox{\tiny A}}(\tilde{x})=v_{\hbox{\tiny P}}(\tilde{x})=\left(1- \frac{\rho_{\hbox{\tiny T}}}{\rho_{\hbox{\tiny A}}(\tilde{x})+\rho_{\hbox{\tiny P}}(\tilde{x})}\right)  \langle V\rangle.
\end{equation}
The observed common velocities at the inner part of the MIPS slab ($\tilde{x}\approx 0$),
$v_{\hbox{\tiny A}}(0)\approx v_{\hbox{\tiny P}}(0)\approx 0.2\approx (1-2/3)\langle V\rangle$, are in good agreement with the total density
profile in Fig. \ref{figure2}, being $\rho_{\hbox{\tiny T}}/(\rho_{\hbox{\tiny A}}(0)+\rho_{\hbox{\tiny P}}(0)) \approx 0.8/1.2=2/3$.

We now demonstrate how the system may develop the asymmetric density profiles, to produce a 
global force $F_{\hbox{\tiny PA}}$ and self-sustain the movement $\langle V\rangle\neq 0$. 
In the ABP model the local rectification of the active force \cite{Chacon2022}
produces the force density $f^{\hbox{\tiny a}}(\tilde{x})$ with positive/negative peaks at
the back/front edges, compressing (i.e. stabilizing) the dense slab.
The null total integral of $f^{\hbox{\tiny a}}(\tilde{x})$ is forced, on average, by the 
symmetry of the density profile. 
However, there are fluctuations  
such that $f^{\hbox{\tiny a}}(\tilde{x})$ becomes stronger at one edge, in correspondence to   
a sharper rise in $\rho_{\hbox{\tiny A}}(\tilde{x})$. Such fluctuation pushes the dense band 
in one direction and the front edge would sweep particles from the low-density phase.
In the pure $\eta_{\hbox{\tiny P}}=0$ case, 
the addition of more particles at the advancing edge would make stronger the rectification of the active force, thus compensating the initial asymmetry.
The short persistence time 
$\tau_p\approx 1.8$ %(see figure 3 of SM) 
reflects that kind of structural fluctuations, also observed  in our binary mixtures.

In a mixture, the sweeping effect at the advancing edge would be stronger for the (lower mobility) passive particles. 
That would decrease the local concentration of the (higher mobility) active particles and the (backwards) rectified active force at the advancing edge, enhancing the initial force/density unbalance.
Thus, the global motion of the slab may increase, against the friction of the Brownian dynamics.

%With a typical time $\tau_o(\eta_P)$, it may appear a fluctuation that is strong enough to drive the system into 

At a given passive particle concentration $\eta_P$, a fluctuation might appear in the system  with a typical time $\tau_o(\eta_P)$: this fluctuations could be strong enough to drive the system into a steady-state motion, with asymmetric density profiles and (action/reaction) opposite global forces on the two species.
Eventually, with a typical time $\tau_{\hbox{\tiny SMP}}(\eta_P) \gg \tau_p$, another large fluctuation could break 
the self-sustained unbalance, leading to the end of the SMP.

\section{Size effects analysis}

\textcolor{black}{The above} analysis underlines  the crucial role of  passive particles as a reactive-bath in which the local rectification of the active forces may produce not only  MIPS, but also the self-sustained asymmetry and the steady motion of the dense band.
Since the unbalanced rectification of $f_{\hbox{\tiny A}}(\tilde{x})$ at the edges of the slab has to produce the real motion of all  particles within, 
%we report in SM that (considering a fixed global density) 
the SMP are strongly size-dependent. 

\begin{figure}[h!]
\includegraphics[width=\linewidth]
{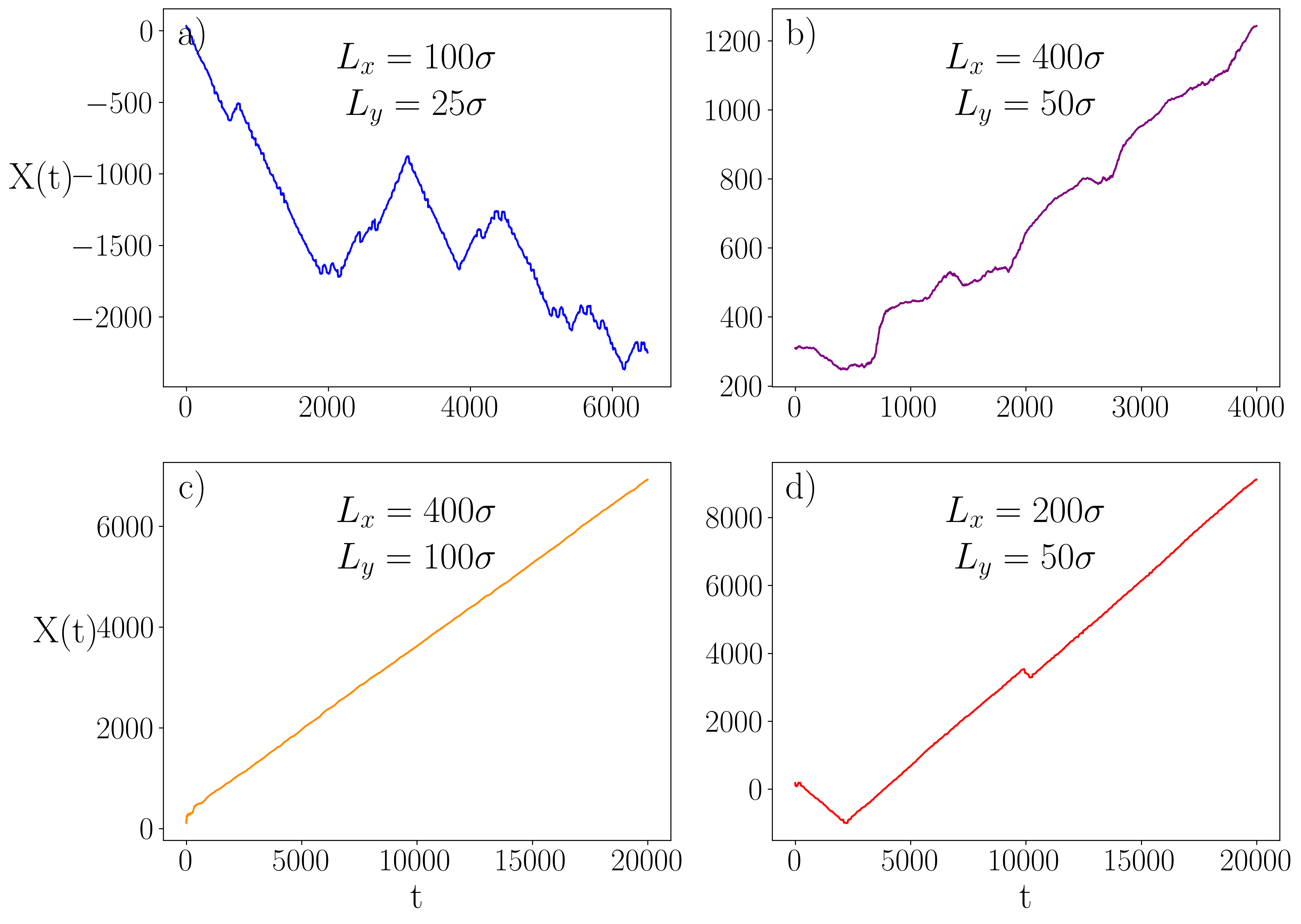}
\caption{\textcolor{black}{Trajectories of MIPS bands ($X(t)$) with $\eta_P=0.4$ for systems of different sizes at a fixed density $\rho = 0.8$ are shown. The trajectories correspond to the following system sizes: panel a, blue, $L_x = 100\sigma$ and $L_y = 25\sigma$ ($N_T=2070$);  panel b, purple, $L_x = 400\sigma$ and $L_y = 50\sigma$ ($N_T=16110$); panel c, orange, $L_x = 400\sigma$ and $L_y = 100\sigma$ ($N_T=32220$); and panel d, red, $L_x = 200\sigma$ and $L_y = 50\sigma$ ($N_T=8055$), the same simulation as in panel d) of Figure \ref{figure1bis}.
}}
\label{Fig_X(t)_sizes}
\end{figure}

\textcolor{black}{Figure~\ref{Fig_X(t)_sizes}, presents  $X(t)$ for several simulations characterised by  $\eta_P = 0.4$ and $\rho_T=0.8$, but different sizes: $L_x = 100\sigma$ and $L_y = 25\sigma$ (a, blue); $L_x = 400\sigma$ and $L_y = 50\sigma$ (b, purple); $L_x = 400\sigma$ and $L_y = 100\sigma$ (c, orange); and $L_x = 200\sigma$ and $L_y = 50\sigma$ (d, red), which corresponds to the size used in all  previous figures.}%, i.e. $X(t)$ in panel d) of Figure \ref{figure1bis} \textcolor{red}{o correr otra simulacion...}.

\textcolor{black}{
The smallest system (in blue, panel (a)) has the box lengths $(L_x,L_y)$ 
and also the slab thickness $(L_s)$, reduced to a half of their values in the (original) system in panel (d).}
\textcolor{black}{
The $X(t)$ trajectory in the smallest system shows SMP as clear as in the original (larger) system, but with a ($\pm$) velocity $V_{\hbox{\tiny SMP}}\approx 1.04$, i.e. much faster that the value $V_{\hbox{\tiny SMP}}\approx 0.57$ of the original system for the same $\eta_P$. That is expected since the total force unbalance has to scale as $L_y$, since
it comes from the interfaces.
On the contrary,  the friction of the Brownian bath has to scale as the total number of particles $N_T=0.8 \ L_x L_y$.  Notice that, even taking into account that the results of the smallest system are reported   over a longer trajectory than those of the larger one, its
typical times seem to be shorter 
(since we observe several U-turns instead
of the single U-turn  observed in the large system).
%simulation of the original system.
A quantitative estimate of the size dependence of $\tau_{\hbox{\tiny SMP}}$ and $\tau_o$ would require extremely large simulations, to get a good statistical sampling of these rare events. However, we might expect longer time scales in larger systems, since they need a larger fluctuation to create or destroy a SMP.
}

\textcolor{black}{
The other two panels in Fig.~\ref{Fig_X(t)_sizes} show the results of simulations in systems larger than the original  one (reported in panel d).  In panels (b) and (c) $L_x$ and the slab thickness $L_s$ are twice as large as in (d). In panel (b) (purple line) we keep the original value of $L_y$, to get a more elongated aspect ratio for the MIPS band. We observe 
that after a long equilibration time the slab takes a positive velocity with $\langle V(t)\rangle\approx 0.3$,
i.e. about half the value of the SMP in the original system (d): this is
consistent with having similar interfacial forces ($\sim L_y$) against double
friction effects ($\sim N_{\hbox{\tiny T}}\sim L_xL_y$). \textcolor{black}{Thus}, the increase of $L_x$ and of the slab thickness $L_s$
leads to  $V(t)$ with much larger fluctuations around its mean (SMP) value, alternating fairly long periods characterised by a 
slab velocity  clearly higher or lower than $0.3$. Thus, the clean separation of the
trajectories in SMP, with $V(t)\approx \pm V_{\hbox{\tiny SMP}}$, divided by static periods with $\langle V(t)\rangle_t\approx 0$,
breaks down.}
\textcolor{black}{This could be caused by the fact that 
%We may interpret that 
the longer $L_x$ and $L_s$ allow for enough free space  
%give room enough 
%to make frequent the presence of 
for the appearance of frequent 
fluctuations that break the slab. Thus,  our description in terms of the single variable $X(t)$  fails.
}

\textcolor{black}{
Panel (c) (orange line) corresponds to a system with the same $L_x$ and slab thickness $L_s$ as in panel (b), but  $L_y$ which is  twice as large as in (d)
(as a consequence, $N_{\hbox{\tiny T}}$ is  four times the original value). The trajectory $X(t)$ has (as in panel (b)) 
a long equilibration period followed by a steady forwards motion of the dense slab, with a mean velocity $\langle V(t)\rangle\approx 0.34$ that
is just slightly higher than in panel (c)  (the system in panel (c) has twice as much interface than in panel (b), and  twice as much $N_{\hbox{\tiny T}}$, i.e. friction from the bath).
 Even though the larger $L_y$ size reduces the fluctuations of $V(t)$
around that mean value, we still cannot  split the trajectory in panel  (c) 
into SMP as cleanly as in the original system. Thus, 
we may interpret that the larger $L_z=100$ makes less likely than with $L_y=50$ that
a fluctuation splits the dense slab into two bands, so that
the description via the variable $X(t)$ (i.e. assuming a single dense band) is more accurate than in panel (b). }

\textcolor{black}{
The size effects in $\langle V\rangle$  are directly transmitted to the mean particles' velocities. Thus,
compared to the values $\langle v_A \rangle = -0.03$ and $\langle v_P \rangle = 0.05$ obtained the original size (panel (d)),
in the larger system (orange panel (c)) with $\langle V \rangle = 0.34$), we obtain that the velocities of each species 
are $\langle v_A \rangle \approx  -0.02$ and $\langle v_P \rangle \approx 0.03$. Very similar values
are obtained when we duplicate $L_x$ keeping the original value of $L_y$.
 Therefore, we conclude that the velocity of the slab is (roughly) inversely proportional to the $x$-length, 
 such that 
$\langle V \rangle \propto \langle v_\alpha \rangle \propto 1 / L_x$
and
$V_{\hbox{\tiny SMP}}\propto \langle v_\alpha\rangle \propto 1/L_x$.
} 
%Moreover, the typical times $\tau_o$ and $\tau_{\hbox{\tiny SMP}}$, for the formation and breakage of SMP, strongly depends  on $L_x$ and $L_y$.

\section{Analysis of U-turn events}

The whole phenomenology described so far has to be interpreted as the appearance of mesoscopic
structures, with large spatial and temporal scales that emerge by the presence of passive particles in
ABP systems. In general, we may expect complex kinetic behaviours consisting of large and fast moving clusters colliding, merging and splitting. 
It is only through our choices for $L_x$ and $L_y$, that we have been able to describe the phenomenology observed using 
%analyse that kinetic behaviour in terms of the SMP, using 
the mesoscopic variable $X(t)$ and the time averages 
with $\tau_p\ll \Delta t < \tau_{\hbox{\tiny SMP}}(\eta_{\hbox{\tiny P}})$. 

\begin{figure}[h!]
\includegraphics[width=\linewidth]{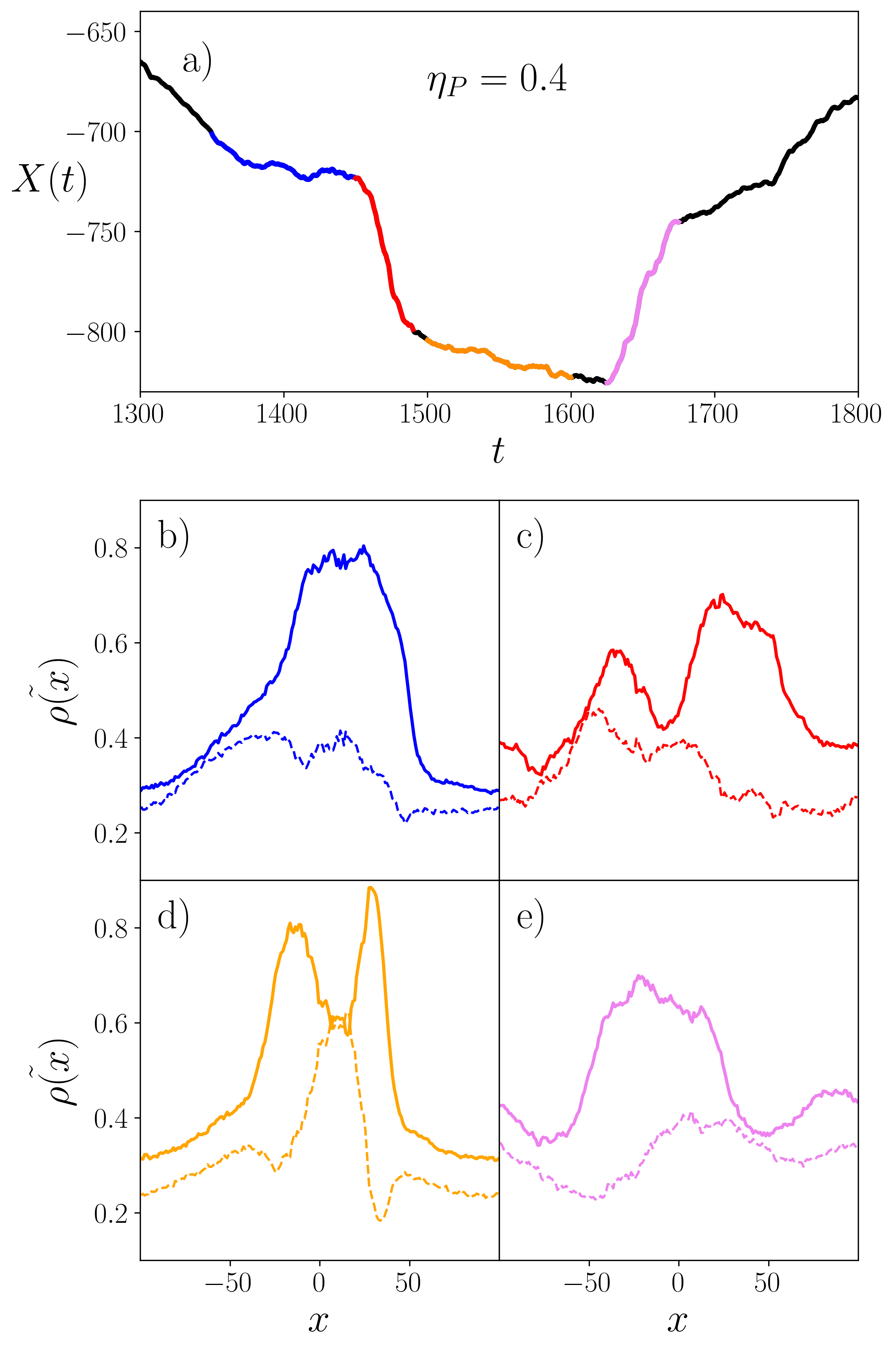}% Here is how to import EPS art
\caption{a) Zoom of the slab centre of mass position ($X(t)$) of the $\eta_{P}=0.4$ mixture in Fig, \ref{figure1bis}-panel d, undergoing a U-turn. Panels b to e show the mean density profiles of the different intermediate states in panel a 
%remarked at the a panel 
between the change of motion.  Active particles density is represented with continuum lines and passive one with dashed lines. The colour code in panels b,c,d,e refers to the colours along the line in panel a.}
\label{figure5}
\end{figure}

The analysis of the events  creating and destroying the SMP has to be restricted 
to shorter time averages and noisier density profiles. 
Figure \ref{figure5}- panel a) represents a zoom-in of the  U-turn observed in Fig.\ref{figure1bis}-panel d at $t\approx 1500$ 
 ($\eta_{\hbox{\tiny P}}=0.4$). 
%The U-turn observed in Fig.\ref{figure1bis}-panel d at $t\approx 1500$  with $\eta_{\hbox{\tiny P}}=0.4$, is amplified Fig.\ref{figure5}- panel a)
(see also video in SM Vid\_SP\_U-turn\_v2.mp4). 

The initial structure, with $V(t)=-V_{\hbox{\tiny SMP}}$,
is characterised by the 
%reserved 
asymmetry reported Fig.\ref{figure2}. Zooming in, panels (b-e) in Fig.\ref{figure5} represent the mean density profiles of active (continuum) and passive (dashed) particles along the U-turn: clearly, the density of the structure is highly disordered,  
%, that structure is strongly disturbed, with the
as shown by the 
appearance of a region with relatively low density of active particles at the middle of the dense slab.

The (nominal) position of the slab $X(t)$, from Eq. (\ref{Eq_Mass_Centre_Fourier}), reflects that a fluctuation  period (nearly flat, in blue and in panel b) is followed by a much faster period (in red and in panel c). 
In fact, the trajectory $X(t)$ itself does not give a  reliable representation  of the actual kinetics. The video in SM$^\dag$ shows that the dense slab is broken into two pieces (panel c) that move in opposite directions until, due to periodic boundaries, they collide again. 

Finally, the system reaches a nearly static structure
over a period of $\Delta t\sim 1000$ (in orange and in panel d), in which the density profiles are (nearly) symmetric, with a  very high density of passive particles at the centre.
Reversing the sequence of events for $t<1500$, 
$X(t)$ has a sudden acceleration (in purple and in panel e) in which the left-right symmetry is broken, to reach a SMP with $V(t)=V_{\hbox{\tiny SMP}}$.

Thus, the large fluctuations %with typical times $\tau_o$ and $\tau_{\hbox{\tiny SMP}}$ 
that create and destroy the SMP in our simulations 
%correspond to mesoscopic  fluctuations that 
overcome our effort, in Eqs. (\ref{Eq_Mass_Centre_Fourier}-\ref{Eq_mean_density_profile}), 
to describe the MIPS structure as a single cluster at a position $X(t)$.

\section{Discussion and Conclusions}

%To conclude, 
\textcolor{black}{In a pure system of active Brownian particles the motion of  particles (induced by the thermal agitation, the active and the interaction forces) is damped by the inert bath, that acts independently on each particle. The addition of repulsive passive particles to ABP systems provides a reactive bath for the active ones. 
Clustering of active particles in a MIPS induces inhomogeneous distributions of the passive particles that would be reflected in the motion of the active ones. Fluctuations that break the 
($\pm x$) symmetry may create total (opposite) forces that  particles of each species induce on those of the other species. Their opposite fluxes on the MIPS less-dense phase  
may lead to a self-sustained motion of the dense band, by a combination of sink-source effects at the edges and a global shift of the (much rigid) inner dense slab. 
}

\textcolor{black}{To characterize and quantify all these effects, we have}
used simulations of a mixture of active/passive Brownian particles in a simple geometry: dense bands that cover the full period $L_y$, along the (shorter) $Y$
axis, and move along the (longer) $X$ axis, so that we may
define a variable $X(t)$ to localize the position of the dense slab and $V(t)$
for its velocity. 
In terms of these mesoscopic variables,  we observe the emergence of characteristic time scales,
much larger than the autocorrelation time $\tau_p$ for $V(t)$
in pure ABP systems. Steady moving structures
(either with positive or  negative mean velocity,
$X(t)-X(t_o)\approx \pm V_{\hbox{\tiny SMP}}(\eta_P) (t-t_o)$)
appear with a typical time $\tau_o(\eta_{\hbox{\tiny P}})$ and 
persist for a typical time $\tau_{\hbox{\tiny SMP}}(\eta_{\hbox{\tiny P}})$ that may be very long. In our simulations we get fairly accurate estimations for $V_{\hbox{\tiny SMP}}(\eta_P)$ in mixtures with different compositions $0.2\leq \eta_P\leq 0.4$, and for different sizes of the simulation box.

\textcolor{black}{The self-assembly and motion of the MIPS clusters is a mesoscopic phenomenon, depending on the size and geometry of the dense cluster and on the distance between its
two edges through the less dense phase (across the periodic
boundaries conditions). The U-turn events observed in our simulations, when the motion of the dense slab is reversed,
correspond to fluctuations that break the slab into two pieces. If the slab is too elongated (large aspect ratio $L_s/L_y$) these events may become frequent, with small pieces of slab breaking out of the main cluster, at one or the other edge, 
with the result of a worse defined velocity $\pm V_{\hbox{\tiny SMP}}$. On the other hand, larger values of $L_y$ give longer 
persistence times $\tau_{\hbox{\tiny SMP}}$, for the steady motion of the slab.}

\textcolor{black}{We compute statistical averages over  (long) time intervals in which $X(t)$ keeps the same linear trend
(either with $V(t)\approx \pm V_{\hbox{\tiny SMP}}(\eta_P)$), 
in terms of particles positions  relative to the slab's 
mesoscopic position  $X(t)$. This allows us to obtain density,
force and velocity profiles  characterizing  kinetics 
and  dynamics of the moving structures. All these variables contribute to the  %that fit well in the 
generic terms of the  Smoluchoski equation for Brownian particles.
}
%With the appropriate choices for $L_x$, $L_y$, $\rho_{\hbox{\tiny T}}$ and $\eta_{\hbox{\tiny P}}$, we
\textcolor{black}{ The generic mechanism to achieve steady motion is that } 
active particles push (and are pushed by)  passive particles, producing stationary density and 
velocity profiles, in terms of the variable $\tilde{x}=x-X(t)$,  reflecting (and producing) a global inter-species 
force $F_{\hbox{\tiny AP}}=-F_{\hbox{\tiny PA}}\neq 0$, either positive or negative.

Thus, the spontaneous breakage of the (spatial $x$ and temporal $t$) symmetry cannot appear in a pure ($\eta_{\hbox{\tiny P}}=0$) ABP system. Thus, we might consider the   passive particles concentration $\eta_{\hbox{\tiny P}}$ as the parameter 
to turn on the most primitive features of hydrodynamics, when collective motions of swimmers are made possible 
by their global push on a reactive medium.
 Our detailed statistical analysis for the different kinetic regimes (steady moving structures and the rare events that produce their U-turns) allowed us to interpret their dynamics within the framework of the Smoluchowski equation
for interacting Brownian particles,  thus providing a clearer physical perspective on  MIPS. 

The dynamics of the MIPS in active-passive mixtures under different conditions is likely to be much complex, with entangled 
effects of the different kinetic regimes  we have identified and analysed. Nevertheless, we  hope that our analysis underlines  the essential role of  passive particles, allowing for the time-space symmetry breaking, over time scales much larger that the typical fluctuations in ABP, and based on a self-sustained combination of a collective drift of  particles in a cluster with local source/sink effects at the edges, leading to the structures' kinetics   apparently faster than the actual particles' mean motion. 
\textcolor{black}{As far as we are aware, so far experimental realizations of MIPS have  dealt with confined systems. However, in order to observe the experimental equivalent of a travelling band, one might need to be able to implement a cylindrical geometry in an experimental set-up.}

\section*{Acknowledgements}
We acknowledge the support of the Spanish
Secretariat for Research, Development and Innovation
(Grants No. PID2020-117080RB-C52, %Pedro tarazona
PID2022-139776NB-C66,   % E. Chacón
IHRC22/00002  
and PID2022-140407NB-C21) 
and from the Maria de Maeztu Programme for Units of Excellence in R\&D
(CEX2023-001316-M). % Pedro Tarazona

%%%END OF MAIN TEXT%%%


\balance


%%%REFERENCES%%%
\bibliography{rsc} %You need to replace "rsc" on this line with the name of your .bib file
\bibliographystyle{rsc} %the RSC's .bst file

\end{document}